\documentclass[10pt,journal,compsoc]{IEEEtran}
\usepackage{hrefhide}
\usepackage{graphicx}
\usepackage{amsfonts}
\usepackage{multirow}
\usepackage{makecell}

\newcolumntype{C}[1]{>{\centering\arraybackslash}p{#1}}

\ifCLASSOPTIONcompsoc
\usepackage[nocompress]{cite}
\else
\usepackage{cite}
\fi
\ifCLASSINFOpdf
\else
\fi
\usepackage{amsmath}
\usepackage{algorithmic}

\usepackage{array}
\usepackage{fixltx2e}

\usepackage{stfloats}
\usepackage[numbers]{natbib}

\usepackage{mathtools}
\DeclarePairedDelimiter{\ceil}{\lceil}{\rceil}
\begin{document}
%
\title{ COVID-19 Cough Classification using Machine Learning and Global Smartphone Recordings }
%
%
%
%

\author{Madhurananda Pahar,
	Marisa Klopper, 
	Robin Warren,
	and Thomas Niesler
	\IEEEcompsocitemizethanks{\IEEEcompsocthanksitem Madhurananda Pahar and Thomas Niesler works at Department of Electrical and Electronics Engineering, University of Stellenbosch, Stellenbosch, South Africa - 7600.\protect\\
		E-mail: mpahar@sun.ac.za, trn@sun.ac.za
		\IEEEcompsocthanksitem Marisa Klopper and Robin Warren works at SAMRC Centre for Tuberculosis Research, University of Stellenbosch, Cape Town, South Africa - 7505.\protect\\
		E-mail: marisat@sun.ac.za, rw1@sun.ac.za
	}
	\thanks{}}

\newcommand{\trn}[1]{\textcolor{red}{[trn: #1]}}
\newcommand{\mk}[1]{\textcolor{purple}{[Marisa: #1]}}

\IEEEtitleabstractindextext{%

\begin{abstract}
	We present a machine learning based COVID-19 cough classifier which can discriminate COVID-19 positive coughs from both COVID-19 negative and healthy coughs recorded on a smartphone. 
	This type of screening is non-contact, easy to apply, and can reduce the workload in testing centres as well as limit transmission by recommending early self-isolation to those who have a cough suggestive of COVID-19.  
	The datasets used in this study include subjects from all six continents and contain both forced and natural coughs, indicating that the approach is widely applicable. 
	The publicly available Coswara dataset contains 92 COVID-19 positive and 1079 healthy subjects, while the second smaller dataset was collected mostly in South Africa and contains 18 COVID-19 positive and 26 COVID-19 negative subjects who have undergone a SARS-CoV laboratory test. 
	Both datasets indicate that COVID-19 positive coughs are 15\%-20\% shorter than non-COVID coughs.
	Dataset skew was addressed by applying the synthetic minority oversampling technique (SMOTE).
	A leave-$p$-out cross-validation scheme was used to train and evaluate seven machine learning classifiers: logistic regression (LR), k-nearest neighbour (KNN), support vector machine (SVM), multilayer perceptron (MLP), convolutional neural network (CNN), long short-term memory (LSTM) and a residual-based neural network architecture (Resnet50). 
	Our results show that although all classifiers were able to identify COVID-19 coughs, the best performance was exhibited by the Resnet50 classifier, which was best able to discriminate between the COVID-19 positive and the healthy coughs with an area under the ROC curve (AUC) of 0.98.
	An LSTM classifier was best able to discriminate between the COVID-19 positive and COVID-19 negative coughs, with an AUC of 0.94 after selecting the best 13 features from a sequential forward selection (SFS). 
	Since this type of cough audio classification is cost-effective and easy to deploy, it is potentially a useful and viable means of non-contact COVID-19 screening. 
\end{abstract}

\begin{IEEEkeywords}
	COVID-19, cough classification, machine learning, logistic regression (LR), k-nearest neighbour (KNN), support vector machine (SVM), multilayer perceptron (MLP), convolutional neural network (CNN), long short-term memory (LSTM), Resnet50
\end{IEEEkeywords}}

\maketitle

\IEEEdisplaynontitleabstractindextext

%
\IEEEpeerreviewmaketitle

\IEEEraisesectionheading{\section{Introduction}\label{sec:introduction}}
\IEEEPARstart{C}{OVID19} (\textbf{CO}rona \textbf{VI}rus \textbf{D}isease of 20\textbf{19}), caused by the Severe Acute Respiratory Syndrome Coronavirus 2 (SARS-CoV2) virus, was declared a global pandemic on February 11, 2020 by the World Health Organisation (WHO). 
It is a new coronavirus but similar to other coronaviruses, including SARS-CoV (severe acute respiratory syndrome coronavirus) and MERS-CoV (Middle East respiratory syndrome coronavirus) which caused disease outbreaks in 2002 and 2012, respectively \cite{world2003summary, miyata2012oxidative}. 

The most common symptoms of COVID-19 are fever, fatigue and dry coughs \cite{wang2020clinical}.
Other symptoms include shortness of breath, joint pain, muscle pain, gastrointestinal symptoms and loss of smell or taste \cite{carfi2020persistent}. 
At the time of writing, there were 142.1 million active cases of COVID-19 globally, and there had been 3 million deaths, with the USA reporting the highest number of cases (31.7 million) and deaths (567,729)~\cite{johnCOVID19}. 
The scale of the pandemic has caused some health systems to be overrun by the need for testing and the management of cases.

Several attempts have been made to identify early symptoms of COVID-19 through the use of artificial intelligence applied to images. 
The 50-layer residual neural network (Resnet50) architecture has been shown to perform better than other pre-trained models such as AlexNet, GoogLeNet and VGG16 in these tasks. 
For example, it has been demonstrated that COVID-19 can be detected from computed tomography (CT) images with an accuracy of 96.23\% by using a Resnet50 architecture \cite{walvekar2020detection}.
The same architecture was shown to detect pneumonia due to COVID-19 with an accuracy of 96.7\%~\cite{sotoudeh2020artificial} and to detect COVID-19 from x-ray images with an accuracy of 96.30\% \cite{yildirim2020deep}. 

Coughing is one of the predominant symptoms of COVID-19~\cite{chang2008chronic} and also a symptom of more than 100 other diseases, and its effect on the respiratory system is known to vary~\cite{higenbottam2002chronic}. 
For example, lung diseases can cause the airway to be either restricted or obstructed and this can influence the acoustics of the cough~\cite{chung2008prevalence}. 
It has also been postulated that the glottis behaves differently under different pathological conditions \cite{korpavs1996analysis, knocikova2008wavelet} and that this makes it possible to distinguish between coughs due to TB \cite{botha2018detection, pahar2021deep}, asthma \cite{al2013signal}, bronchitis and pertussis (whooping cough)~\cite{pramono2016cough, windmon2018tussiswatch, sharan2018automatic, rudraraju2020cough}.

Respiratory data such as breathing, sneezing, speech, eating behaviour and coughing can be processed by machine learning algorithms to diagnose respiratory illness \cite{deshpande2020overview, belkacem2021end, schuller2020covid}. 
Simple machine learning tools, like binary classifiers, are able to distinguish COVID-19 respiratory sounds from healthy counterparts with an area under the ROC curve (AUC) exceeding 0.80~\cite{brown2020exploring}. 
Detecting COVID-19 by analysing only the cough sounds is also possible. 
\mbox{AI4COVID-19} is a mobile app that records 3 seconds of cough audio which is analysed automatically to provide an indication of COVID-19 status within 2 minutes~\cite{imran2020ai4covid}. 
A deep neural network (DNN) was shown to distinguish between COVID-19 and other coughs with an accuracy of 96.83\% on a dataset containing 328 coughs from 150 patients of four different classes: COVID-19, asthma, bronchitis and healthy ~\cite{pal2021pay}. 
There appear to be unique patterns in COVID-19 coughs that allow a pre-trained Resnet18 classifier to identify COVID-19 coughs with an AUC of 0.72.
In this case, cough samples were collected over the phone from 3621 individuals with confirmed COVID-19~\cite{bagad2020cough}.  
COVID-19 coughs were classified with a higher AUC of 0.97 (sensitivity = 98.5\% and specificity = 94.2\%) by a Resnet50 architecture, trained on coughs from 4256 subjects and evaluated on 1064 subjects that included both COVID-19 positive and COVID-19 negative subjects by implementing four biomarkers \cite{laguarta2020covid}. 
A high AUC exceeding 0.98 was also achieved in \cite{andreu2021generic} when discriminating COVID-19 positive coughs from COVID-19 negative coughs on a clinically validated dataset consisting of 2339 COVID-19 positive and 6041 COVID-19 negative subjects using DNN based classifiers. 

Data collection from COVID-19 patients is challenging and the datasets are often not publicly available.
Nevertheless, efforts have been made to compile such datasets. 
For example, a dataset consisting of coughing sounds recorded during or after the acute phase of COVID-19 from patients via public media interviews has been developed in~\cite{cohen2020novel}. 
The Coswara dataset is publicly available and collected in a more controlled and targeted manner~\cite{sharma2020coswara}.
At the time of writing, this dataset included usable `deep cough' i.e. loud coughs recordings from 92 COVID-19 positive and 1079 healthy subjects.
We have also begun to compile our own dataset by collecting recordings from subjects who have undergone a SARS-CoV laboratory test in South Africa. 
This Sarcos (\textbf{SAR}S \textbf{CO}VID-19 \textbf{S}outh Africa) dataset is currently still small and includes only 44 subjects (18 COVID-19 positive and 26 COVID-19 negative).

\begin{figure*}[ht!]
	\centerline{\includegraphics[width=0.99\textwidth]{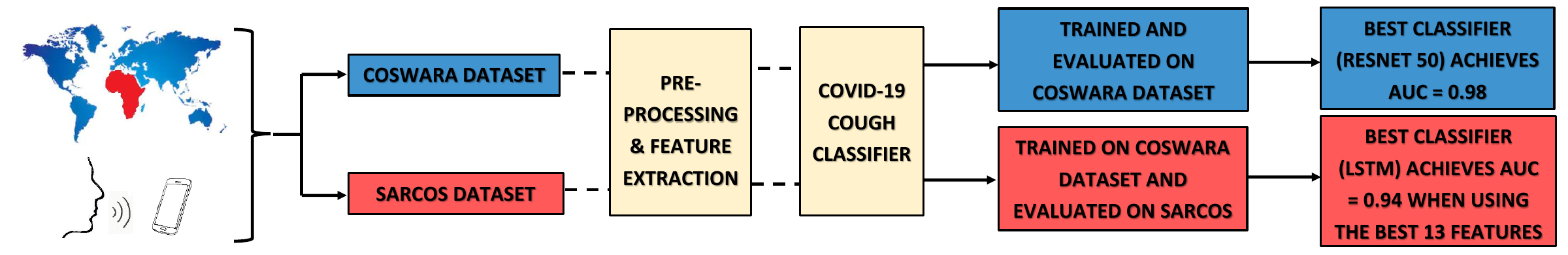}}
	\caption{\textbf{Location of participants in the Coswara and the Sarcos datasets:} Participants in the Coswara dataset were located on five different continents, excluding Africa. The majority (91\%) of participants in the Coswara dataset are from Asia, as indicated in Figure \ref{fig:coswara-dataset}. Sarcos participants who supplied geographical information are mostly (75\%) from South Africa, as shown in Figure \ref{fig:sarcos-dataset}.} 
	\label{fig:class-train-test}
\end{figure*}

Both the Coswara and Sarcos datasets are imbalanced since COVID-19 positive subjects are outnumbered by non-COVID-19 subjects. 
Nevertheless, collectively these two datasets contain recordings from all six continents, as shown in Figure \ref{fig:class-train-test}. 
To improve machine learning classification performance, we have applied the synthetic minority over-sampling technique (SMOTE) to balance our datasets. 
Furthermore, we have found that the COVID-19 positive coughs are 15\%-20\% shorter than non-COVID coughs.
Hence, feature extraction is designed to preserve the time-domain patterns over an entire cough. 
Classifier hyperparameters were optimised by using the leave-$p$-out cross-validation, followed by training and evaluation of
machine learning approaches, namely logistic regression (LR), k-nearest neighbour (KNN), support vector machine (SVM), multilayer perceptron (MLP) and deep neural networks (DNN) such as a convolutional neural network (CNN), long short-term memory (LSTM) and Resnet50. 
The Resnet50 produced the highest AUC of 0.976 $\approx$ 0.98 when trained and evaluated on the Coswara dataset, outperforming the baseline results presented in \cite{muguli2021dicova}. 
No classifier has been trained on the Sarcos dataset due to its small size.
It can also not be combined with Coswara as it contains slightly different classes. 
Instead, this dataset has been used for an independent validation of the best-performing DNN classifiers developed on the Coswara dataset. 
In these validation experiments, it was found that the highest AUC of 0.938 $\approx$ 0.94 is achieved when using the best 13 features identified using the greedy sequential forward selection (SFS) algorithm and an LSTM classifier. 
We conclude that it is possible to identify COVID-19 on the basis of cough audio recorded using a smartphone.
Furthermore, this discrimination between COVID-19 positive and both COVID-19 negative and healthy coughs is possible for audio samples collected from subjects located all over the world. 
Additional validation is however still required to obtain approval from regulatory bodies for use as a diagnostic tool. 


\section{Data}
We have used two datasets in our experimental evaluation: the Coswara dataset and the Sarcos dataset.

\subsection{The Coswara Dataset}
\begin{figure}[h!]
	\centerline{\includegraphics[width=0.5\textwidth]{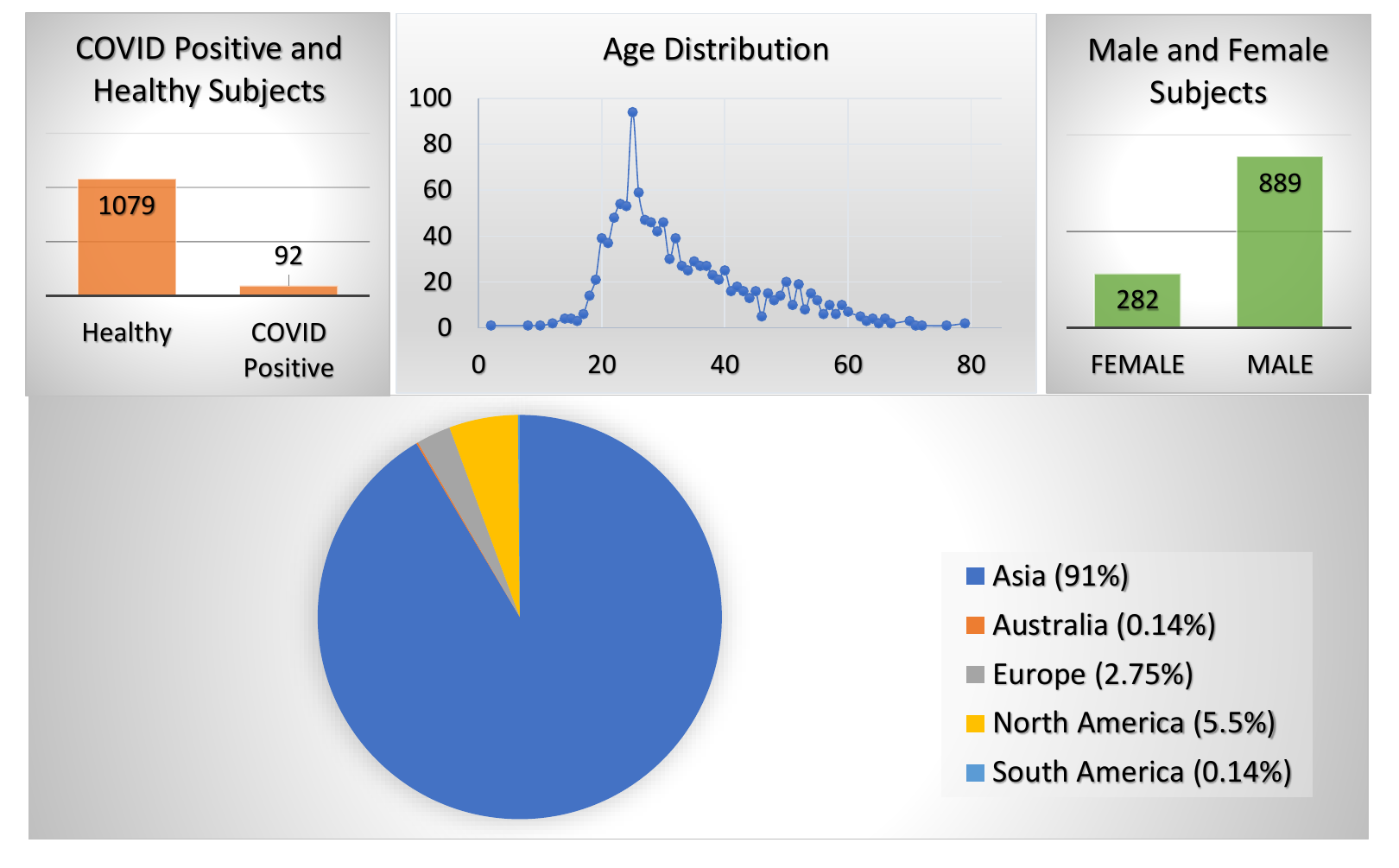}}
	\caption{\textbf{Coswara dataset at the time of experimentation:} There are 1079 healthy and 92 COVID-19 positive subjects in the pre-processed dataset, used for feature extraction and classifier training. Most of the subjects are aged between 20 and 50. There are 282 female and 889 male subjects and most of them are from Asia. Subjects are from five continents: \textbf{Asia} (Bahrain, Bangladesh, China, India, Indonesia, Iran, Japan, Malaysia, Oman, Philippines, Qatar, Saudi Arabia, Singapore, Sri Lanka, United Arab Emirates), \textbf{Australia}, \textbf{Europe} (Belgium, Finland, France, Germany, Ireland, Netherlands, Norway, Romania, Spain, Sweden, Switzerland, Ukraine, United Kingdom), \textbf{North America} (Canada, United States), and \textbf{South America} (Argentina, Mexico).}
	\label{fig:coswara-dataset}
\end{figure}

The Coswara project is aimed at developing a diagnostic tool for COVID-19 based on respiratory, cough and speech sounds~\cite{sharma2020coswara}. 
Public participants were asked to contribute cough recordings via a web-based data collection platform using their smartphones (\url{https://coswara.iisc.ac.in}). 
The collected audio data includes fast and slow breathing, deep and shallow coughing, phonation of sustained vowels and spoken digits. 
Age, gender, geographical location, current health status and pre-existing medical conditions are also recorded. 
Health status includes `healthy', `exposed', `cured' or `infected'. 
Audio recordings were sampled at 44.1 KHz and subjects were from all continents except Africa, as shown in Figure \ref{fig:coswara-dataset}.
In this study, we have made use of the raw audio recordings and applied pre-processing as described in Section \ref{subsec:preprocess}.

\subsection{The Sarcos Dataset}
A similar initiative in South Africa encouraged participants to allow the voluntarily recording their coughs  using an online platform (\url{https://coughtest.online}) under the research project name: `COVID-19 screening by cough sound analysis'. 
This dataset will be referred to as `Sarcos' (\textbf{SAR CO}VID-19 \textbf{S}outh Africa). 
Only coughs were collected as audio samples, and only subjects who had recently undergone a SARS-CoV laboratory test were asked to participate. 
The sampling rate for the audio recordings was 44.1 KHz. 
In addition to the cough audio recordings, subjects were presented with a voluntary and anonymous questionnaire, providing informed consent. 
The questionnaire prompted for the following information. 

\begin{itemize}
	\item Age and gender.
	\item Whether tested by an authorised COVID-19 testing centre.
	\item Days since the test was performed.
	\item Lab result (COVID-19 positive or negative).
	\item Country of residence. 
	\item Known contact with COVID-19 positive patient.
	\item Known lung disease.
	\item Symptoms and temperature. 
	\item Whether they are a regular smoker. 
	\item Whether they have a current cough and for how many days. 
\end{itemize}

\begin{figure}[h!]
	\centerline{\includegraphics[width=0.5\textwidth]{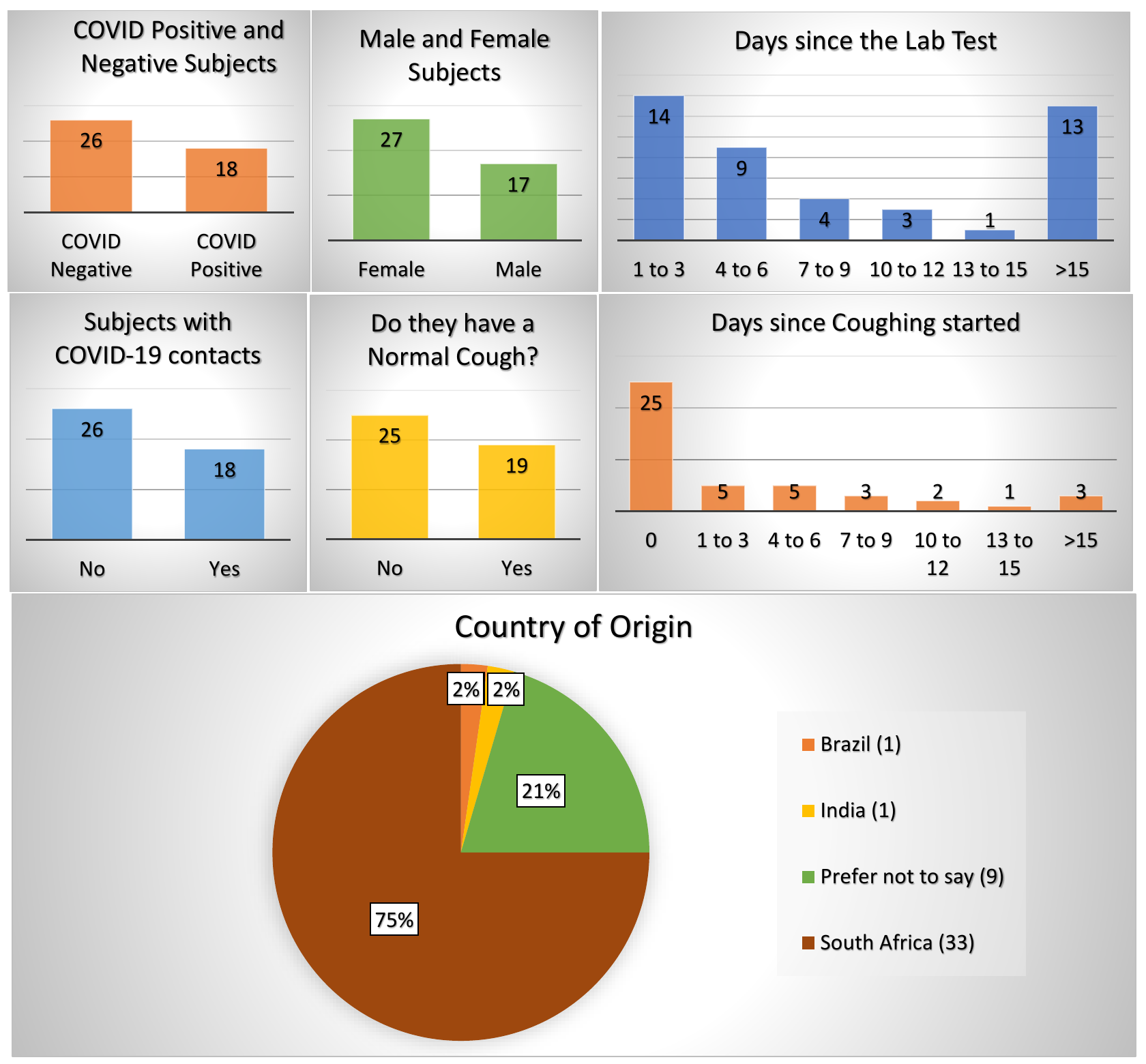}}
	\caption{\textbf{Sarcos dataset at the time of experimentation:} There are 26 COVID-19 negative and 18 COVID-19 positive subjects in the processed dataset. Unlike the Coswara dataset, there are more female than male subjects. Most of the subjects had their lab test performed within two weeks of participation. 
		Only 19 of the subjects reported coughing as a symptom, and for these the reported duration of coughing symptoms was variable. There were 33 subjects from \textbf{Africa} (South Africa),  1 from \textbf{South America} (Brazil), 1 from \textbf{Asia} (India) and the rest declined to specify their geographic location.}
	\label{fig:sarcos-dataset}
\end{figure}

Among the 44 participants, 33 (75\%) subjects asserted that they are South African residents and therefore represent the African continent, as shown in Figure \ref{fig:sarcos-dataset}. 
As there were no subjects from Africa in the Coswara dataset, together the Coswara and Sarcos dataset include subjects from all six continents. 

\subsection{Data Pre-processing}
\label{subsec:preprocess}
The raw cough audio recordings from both datasets have the sampling rate ($\mu$) of $44.1$ KHz and
is subjected to some simple pre-processing steps, described below. 
We note, 
time-window length ($\lambda$) as $0.05$ seconds and amplitude threshold value ($\Phi$) as $0.005$,
where both of these values were determined manually and interactively, as the silence removal was validated by visual inspection in all cases.

%
%
The original cough audio $c_i(t)$ is normalised by following Equation \ref{eq:norm}.

\begin{equation} \label{eq:norm}
	c_i(t) = 0.9 \times \frac{c_i(t)}{\lvert max(c_i(t)) \rvert}
\end{equation}

\begin{figure}[h!]
	\centerline{\includegraphics[width=0.5\textwidth]{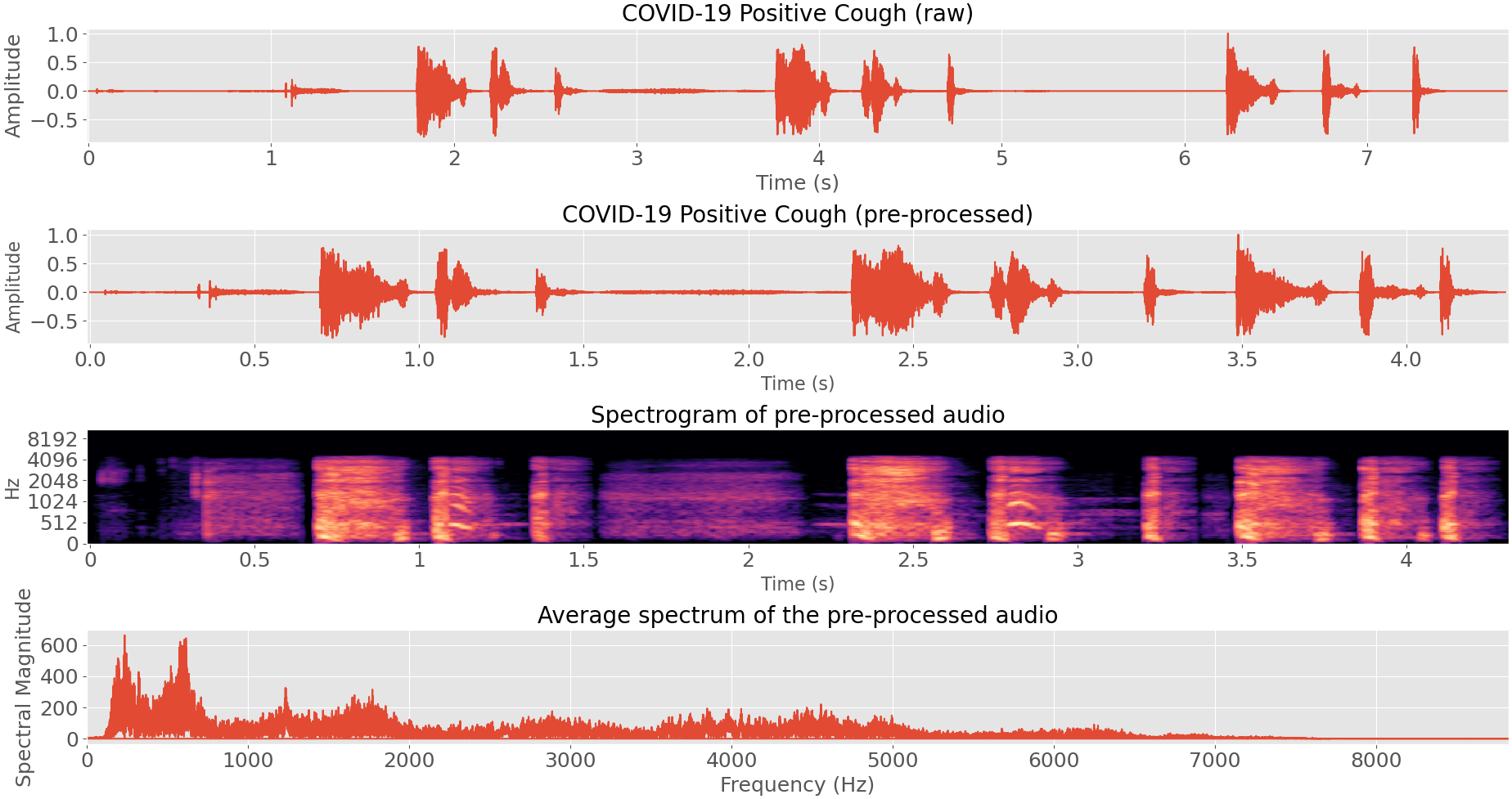}}
	\caption{\textbf{A processed COVID-19 cough audio} which is shorter than the original cough audio but keeps all spectrum resolution. Amplitudes are normalised and extended silences are removed in the pre-processing. }
	\label{fig:processed-cough-info}
\end{figure}

The processed final cough audio is shown in Figure \ref{fig:processed-cough-info} and noted as: $C(t)$. 
Here, $I$ denotes the time-window and we define: 
\begin{equation} \label{eq:signal-part}
	C_I(t) = C_{j\mu \lambda}(t) \cdots C_{(j+1)\mu \lambda}(t)
\end{equation}

For example, when $j$ = 0; $C_I$ will be the portion of signal where $C_0 \cdots C_{2205}$, as $\mu =$ 44100 Hz and $\lambda =$ 0.05 seconds. $0 \leq j \leq \lfloor{\frac{\Xi}{\mu \lambda}\rfloor} $, where $\Xi$ is the length of signal $c_i(t)$.
$C(t)$ is calculated by following Equation \ref{eq:processed-signal}. 

\begin{equation} \label{eq:processed-signal}
	C(t) = C(t) \oplus C_I 
	\begin{cases}
		\text{ if } C_I(t) \geq \Phi
	\end{cases}
\end{equation}
where, $\oplus$ means concatenation and, $C_I(t) \geq \Phi$, if $C_{I_i}(t) \geq \Phi $, where $\forall i \in I $. 

Thus, the amplitudes of the raw audio data in the Coswara and the Sarcos dataset were normalised, after which periods of silence were removed from the signal to within a 50 ms margin using a simple energy detector. Figure \ref{fig:processed-cough-info} shows an example of the original raw audio, as well as the pre-processed audio.

After pre-processing, the Coswara dataset contains 92 COVID-19 positive and 1079 healthy subjects and the Sarcos dataset contains 18 COVID-19 positive and 26 COVID-19 negative subjects, as summarised in Table~\ref{table:coswara-sarcos-dataset}.
In both datasets, COVID-19 positive coughs are 15\%-20\% shorter than non-COVID coughs.

\begin{table*}[h]
	\caption{\textbf{Summary of the Coswara and Sarcos Datasets:} In the COSWARA dataset, there were 1171 subjects with usable `deep cough' recordings, 92 of whom were COVID-19 positive while 1079 were healthy. This amounts to a total of 1.05 hours of cough audio recordings (after pre-processing) that will be used for experimentation. The Sarcos dataset contains data from a total of 44 subjects, 18 of whom are COVID-19 positive and 26 who are not. This amounts to a total of 2.45 minutes of cough audio recordings (after pre-processing) that has been used for experimentation. COVID-19 positive coughs are 15\%-20\% shorter than non-COVID coughs. } 
	\centering 
	\begin{center}
		\begin{tabular}{ c | c | c | c | c | c }
			\hline
			\hline
			\textbf{Dataset} & \textbf{Label} & \textbf{Subjects} & \textbf{Total audio} & \textbf{Average per subject } & \textbf{Standard deviation }  \\
			\hline
			\hline
			Coswara & COVID-19 Positive & 92 & 4.24 mins & 2.77 sec & 1.62 sec \\
			\hline
			Coswara & Healthy & 1079 & 0.98 hours & 3.26 sec & 1.66 sec  \\
			\hline
			Coswara & \textbf{Total} & 1171 & 1.05 hours & 3.22 sec & 1.67 sec  \\
			\hline
			\hline
			Sarcos & COVID-19 Positive & 18  & 0.87 mins & 2.91 sec & 2.23 sec \\
			\hline 
			Sarcos & COVID-19 Negative & 26 & 1.57 mins & 3.63 sec & 2.75 sec \\
			\hline
			Sarcos & \textbf{Total} & 44 & 2.45 mins & 3.34 sec & 2.53 sec \\
			\hline
			\hline
		\end{tabular}
	\end{center}
	\label{table:coswara-sarcos-dataset}
\end{table*}

\subsection{Dataset Balancing}
Table~\ref{table:coswara-sarcos-dataset} shows that COVID-19 positive subjects are under-represented in both datasets. 
To compensate for this imbalance, which can detrimentally affect machine learning~\cite{van2007experimental,krawczyk2016learning}, we have applied SMOTE data balancing to create equal number of COVID-19 positive coughs during training~\cite{chawla2002smote,lemaitre2017imbalanced}. 
This technique has previously been successfully applied to cough detection and classification based on audio recordings \cite{windmon2018tussiswatch, pahar2021covidbreath, pahar2021deep}. 

SMOTE oversamples the minor class by generating synthetic examples, instead of for example random oversampling. 
In our dataset, for each COVID-19 positive cough, 5 other COVID-19 positive coughs were randomly chosen and the closest in terms of the Euclidean distance is identified as ${\bf x}^{NN}$. 
Then the synthetic COVID-19 positive samples are created using Equation \ref{eq:SMOTE}. 

\begin{equation} \label{eq:SMOTE}
	{\bf x}^{SMOTE}={\bf x}+u\cdot({\bf x}^{NN}-{\bf x})
\end{equation}

The multiplicative factor $u$ is uniformly distributed between 0 and 1 \cite{BlagusSMOTE}. 

We have also implemented other extensions of SMOTE such as borderline-SMOTE \cite{han2005borderline, nguyen2011borderline} and adaptive synthetic sampling \cite{he2008adasyn}. However, the best results were obtained by using SMOTE without any modification.

\section{Feature Extraction}
\label{sec:feat-extract}

\begin{figure}
	\centerline{\includegraphics[width=0.5\textwidth]{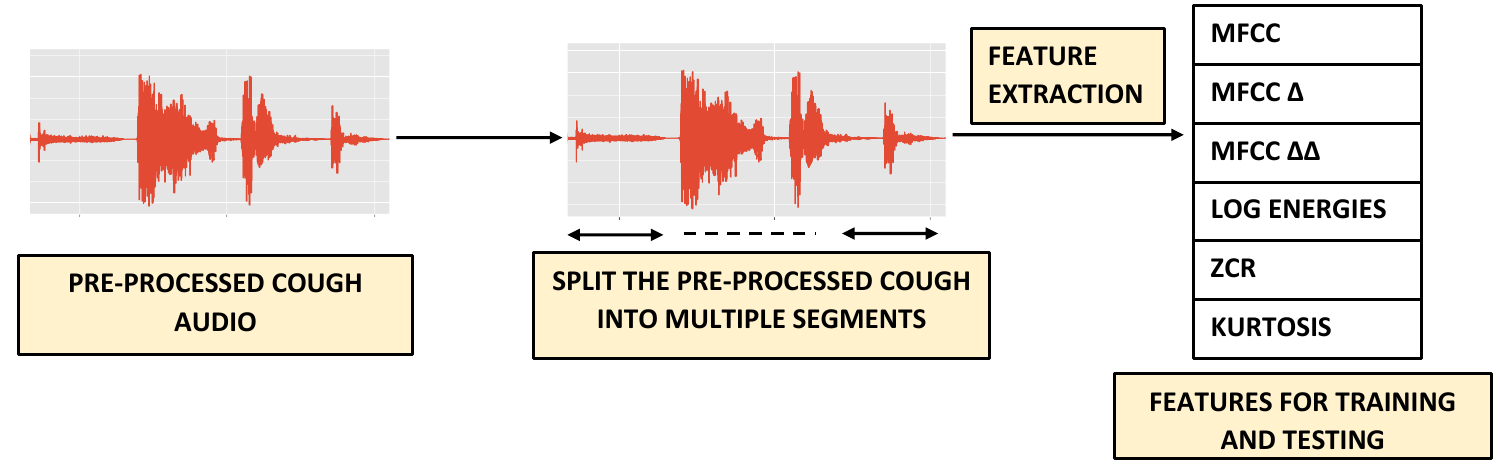}}
	\caption{\textbf{Feature Extraction:} Pre-processed cough audio recordings, shown in Figure \ref{fig:processed-cough-info}, are split into individual segments after which features such as MFCCs, MFCCs velocity ($\Delta$), MFCCs acceleration ($\Delta \Delta$), log frame energies, ZCR and kurtosis are extracted. So, for $\mathcal{M}$ number of MFCCs and $\mathcal{S}$ number of segments, the final feature matrix has ($3\mathcal{M} + 3, \mathcal{S}$) dimensions. }
	\label{fig:feat-extract}
\end{figure}

The feature extraction process is illustrated in Figure~\ref{fig:feat-extract}. 
Features such as mel-frequency cepstral coefficients (MFCCs), log frame energies, zero crossing rate (ZCR) and kurtosis are extracted. 
MFCCs have been used very successfully as features in audio analysis and especially in automatic speech recognition~\cite{WeiHan2006, pahar_coding_2020}.
They have also been found to be useful in differentiating dry coughs from wet coughs \cite{chatrzarrin2011feature} and classifying tuberculosis coughs \cite{pahar2021tb}. 
We have used the traditional MFCC extraction method considering higher resolution MFCCs along with the velocity (first-order difference, $\Delta$) and acceleration (second-order difference, $\Delta \Delta$) as adding these has shown classifier improvement in the past \cite{azmy2017feature}. 
Log frame energies can improve the performance in audio classification tasks \cite{aydin2009log}.
The ZCR \cite{bachu2010voiced} is the number of times a signal changes sign within a frame, indicating the variability present in the signal.
The kurtosis \cite{decarlo1997meaning} indicates the tailedness of a probability density.
For the samples of an audio signal, it indicates the prevalence of higher amplitudes. 
These features have been extracted by using the hyperparameters described in Table~\ref{table:feat-hyper-parameter} for all cough recordings.

We have extracted features in a way that preserves the information regarding the beginning and the end of a cough event to allow time-domain patterns in the recordings to be discovered while maintaining the fixed input dimensionality expected by, for example, a CNN.
From every recording, we extract a fixed number of features $\mathcal{S}$ by distributing the fixed-length analysis frames uniformly over the time-interval of the cough.
The input feature matrix for the classifiers then always has the dimension of ($3\mathcal{M} + 3, \mathcal{S}$) for $\mathcal{M}$ number of MFCCs along with $\mathcal{M}$ number of velocity ($\Delta$) and $\mathcal{M}$ number of acceleration ($\Delta \Delta$), as illustrated in Figure \ref{fig:feat-extract}. 
If $\Lambda$ is the number of samples in the cough audio, we can calculate the number of samples between consecutive frames $\delta$ using Equation \ref{eq:seg}. 

\begin{equation} \label{eq:seg}
\delta = \ceil[\bigg]{ \frac{\Lambda}{\mathcal{S}} }
\end{equation}

So, for example a 2.2 second long cough audio event contains 97020 samples, as the sampling rate is 44.1 KHz. 
If the frame length is 1024 samples and number of segments are 100, then the frame skip ($\delta$) is $\ceil[\bigg]{ \frac{97020}{100} }$ = 971 samples.

In contrast with the more conventionally applied fixed frame rates, this way of extracting features ensures that the entire recording is captured within a fixed number of frames, allowing especially the CNN classifiers to discover more useful temporal patterns and provide better classification performance. 
This particular method of feature extraction has also shown promising result in classifying COVID-19 breath and speech \cite{pahar2021covidbreath}. 

\section{Classifier architectures}

We have trained and evaluated seven machine learning classifiers in total. 
LR models have been found to outperform other more complex classifiers such as classification trees, random forests, SVM in some clinical prediction tasks~\cite{christodoulou2019systematic, botha2018detection, le1992ridge}. 
%
%
We have used gradient descent weight regularisation as well as lasso ($l1$ penalty) and ridge ($l2$ penalty) estimators during training~\cite{tsuruoka2009stochastic, yamashita2003interior}.
%
This LR classifier has been intended primarily as a baseline against which any improvements offered by the more complex architectures can be measured.
A KNN classifier bases its decision on the class labels of the $k$ nearest neighbours in the training set and in the past has been able to both detect \cite{monge2018robust,pramono2019automatic,vhaduri2019nocturnal} and classify \cite{wang2006environmental,pramono2016cough, pahar2021tb} sounds such as coughs and snores successfully. 
SVM classifiers have also performed well in both detecting \cite{bhateja2019pre, tracey2011cough} and classifying \cite{sharan2017cough} cough events. 
%
%
%
%
%
The independent term in kernel functions is chosen as a hyperparameter while optimising the SVM classifier. 
%
%
%
%
An MLP, a neural network with multiple layers of neurons separating the input and output \cite{taud2018multilayer}, is capable of learning non-linear relationships and have for example been shown to be effective when discriminating influenza coughs from other coughs~\cite{sarangi2016design}. 
MLP have also been applied to classify tuberculosis coughs~\cite{tracey2011cough, pahar2021tb} and detect coughs in general~ \cite{liu2014cough, amoh2015deepcough}. 
%
The penalty ratios, along with the number of neurons are used as the hyperparameters which were optimised using the leave-$p$-out cross-validation process (Figure~\ref{fig:lpo-k-fold} and Section \ref{sec:crossvalidation}).

\begin{figure}
	\centerline{\includegraphics[width=0.5\textwidth]{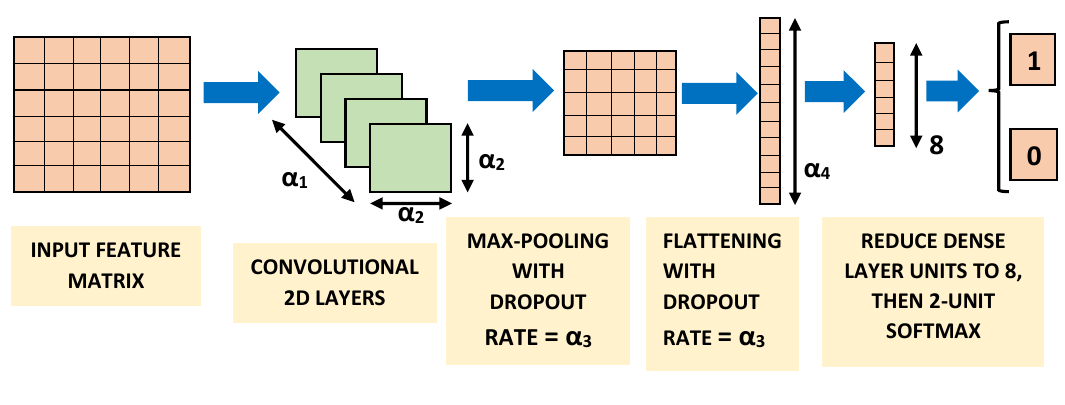}}
	\caption{\textbf{CNN Classifier:} Our CNN classifier uses $\alpha_1$ two-dimansional convolutional layers with kernel size $\alpha_2$, rectified linear units as activation functions and a dropout rate of $\alpha_3$. After max-pooling, two dense layers with $\alpha_4$ and 8 units respectively and rectified linear activation functions follow. The network is terminated by a two-dimensional softmax where one output (1) represents the COVID-19 positive class and the other (0) healthy or COVID-19 negative class. During training, features are presented to the neural network in batches of size $\beta_3$ for $\beta_4$ epochs. }
	\label{fig:CNN-fig}
\end{figure}
\begin{figure}
	\centerline{\includegraphics[width=0.5\textwidth]{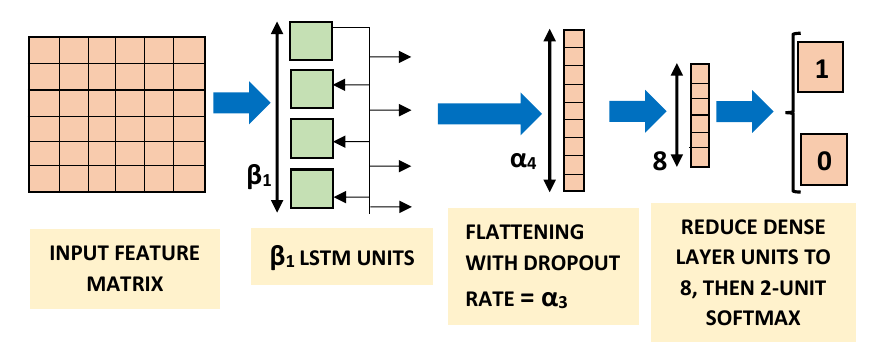}}
	\caption{\textbf{LSTM classifier:} Our LSTM classifier has $\beta_1$ LSTM units, each with rectified linear activation functions and a dropout rate of $\alpha_3$. This is followed by two dense layers with $\alpha_4$ and 8 units respectively and rectified linear activation functions. The network is terminated by a two-dimensional softmax where one output (1) represents the COVID-19 positive class and the other (0) healthy or COVID-19 negative class. During training, features are presented to the neural network in batches of size $\beta_3$ for $\beta_4$ epochs. }
	\label{fig:RNN-fig}
\end{figure}

A CNN is a popular deep neural network architecture, primarily used in image classification~\cite{krizhevsky2017imagenet}. 
For example, in the past two decades CNNs were applied successfully to complex tasks such as face recognition~\cite{lawrence1997face}. 
It has also performed well in classifying COVID-19 breath and speech \cite{pahar2021covidbreath}. 
A CNN architecture \cite{albawi2017understanding, qi2017comparison} along with the optimised hyperparameters (Table \ref{table:class-hyper-parameter}) is shown in Figure \ref{fig:CNN-fig}.
%
%
%
%
%
%
%
%
%
%
An LSTM model is a type of recurrent neural network whose architecture allows it to remember previously-seen inputs when making its classification decision~\cite{hochreiter1997long}. 
It has been successfully used in automatic cough detection~\cite{miranda2019comparative, pahar2021deep}, and also in other types of acoustic event detection~\cite{marchi2015non, amoh2016deep}. 
The hyperparameters optimised for the LSTM classifier \cite{sherstinsky2020fundamentals} are mentioned in Table \ref{table:class-hyper-parameter} and visually explained in Figure \ref{fig:RNN-fig}. 
%
%
%
%
The 50-layer deep residual learning (Resnet50) neural network \cite{he2016deep} is a very deep architecture that contains skip layers, and has been found to outperform other very deep architectures such as VGGNet.
It performs particularly well on image classification tasks on the dataset such as ILSVRC, the CIFAR10 dataset and the COCO object detection dataset \cite{lin2014microsoft}. 
Resnet50 has already been used in successfully detecting COVID-19 from CT images \cite{walvekar2020detection}, coughs \cite{laguarta2020covid}, breath, speech \cite{pahar2021covidbreath} and Alzheimer's \cite{laguarta2020longitudinal}. 
%
%
%
%
Due to extreme computation load, we have used the default Resnet50 structure mentioned in Table 1 of \cite{he2016deep}.

\section{Classification Process}


\subsection{Hyperparameter Optimisation}

Both feature extraction and classifier architectures have a number of hyperparameters. They are listed in Table \ref{table:feat-hyper-parameter} and \ref{table:class-hyper-parameter} and were optimised by using a leave-$p$-out cross-validation scheme. 

\begin{table}[h]
	\caption{\textbf{Feature extraction hyperparameters} optimised using the leave-$p$-out cross-validation as described in Section~\ref{sec:crossvalidation} } 
	\centering 
	\begin{center}
		\begin{tabular}{ c | c | c }
			\hline
			\hline
			\textbf{Hyperparameter} & \textbf{Description}     & \textbf{Range} \\
			\hline
			\hline
			\multirow{2}{*}{MFCC ($\mathcal{M}$)}   & Number of lower-order    & $13 \times k_1$, where \\
			& MFCCs to keep            & $k_1=1, 2, 3, 4, 5$ \\ 
			\hline
			\multirow{2}{*}{Frame ($\mathcal{F}$)}  & Frame-size in which      & $2^{k_2}$ where \\
			& audio is segmented       &  $k_2=8, \cdots, 12$\\
			\hline
			\multirow{2}{*}{Seg ($\mathcal{S}$)}    & Number of frames & $10 \times k_3$, where \\
			& extracted from the audio   & $k_3=5, 7, 10, 12, 15$ \\
			\hline
			\hline
		\end{tabular}
	\end{center}
	\label{table:feat-hyper-parameter}
\end{table}

As the sampling rate is 44.1 KHz in both the Coswara and Sarcos dataset, by varying the frame lengths from $2^8$ to $2^{12}$ i.e. 256 to 4096 samples, features are extracted from frames whose duration varies between approximately 5 and 100~ms. 
Different phases in a cough carry important features \cite{chatrzarrin2011feature} and thus each cough has been divided between 50 and 150 segments with steps of 20 to 30, as shown in Figure \ref{fig:feat-extract}. 
By varying the number of lower-order MFCCs to keep (from 13 to 65, with steps of 13), the spectral resolution of the features was varied. 

\begin{table*}[h]
	\caption{\textbf{Classifier hyperparameters},  optimised using the leave-$p$-out cross-validation as described in Section~\ref{sec:crossvalidation} } 
	\centering 
	\begin{center}
		\begin{tabular}{ c | c | c | c  }
			\hline
			\hline
			\textbf{Hyperparameter} & \textbf{Description} & \textbf{Classifier} & \textbf{Range} \\
			\hline
			\hline
			$\nu_1$     & Regularisation strength      & LR         & $10^{i_1}$ where $i_1=-7,-6,\ldots,6,7$ $(10^{-7}$ to $10^{7})$ \\
			\hline
			$\nu_2$     & $l1$ penalty                 & LR         & 0 to 1 in steps of 0.05 \\
			\hline
			$\nu_3$     & $l2$ penalty                 & LR         & 0 to 1 in steps of 0.05 \\

			\hline
			$\xi_1$    & Number of neighbours      & KNN         & 10 to 100 in steps of 10 \\
			\hline
			$\xi_2$    & Leaf size     & KNN         & 5 to 30 in steps of 5 \\

			\hline
			$\zeta_{1}$    & Regularisation strength      & SVM         & $10^{i_3}$ where $i_3=-7,-6,\ldots,6,7$ $(10^{-7}$ to $10^{7})$ \\
			\hline
			$\zeta_{2}$    & Kernel Coefficient     & SVM         & $10^{i_4}$ where $i_4=-7,-6,\ldots,6,7$ $(10^{-7}$ to $10^{7})$ \\

			\hline
			$\eta_{1}$      & No. of neurons         & MLP        & 10 to 100 in steps of 10 \\
			\hline
			$\eta_{2}$ & $l2$ penalty                 & MLP        & $10^{i_2}$ where $i_2=-7,-6,\ldots,6,7$ $(10^{-7}$ to $10^{7})$ \\
			\hline
			$\eta_{3}$ & Stochastic gradient descent  & MLP        & 0 to 1 in steps of 0.05 \\

			\hline
			$\alpha_1$  & No. of Conv filters          & CNN        & $3 \times 2^{k_4}$ where $k_4=3, 4, 5$ \\
			\hline
			$\alpha_2$  & Kernel size                 & CNN        & 2 and 3 \\
			\hline
			$\alpha_3$  & Dropout rate                & CNN, LSTM  & 0.1 to 0.5 in steps of 0.2 \\
			\hline
			$\alpha_4$  & Dense layer size             & CNN, LSTM  & $2^{k_5}$ where $k_5=4, 5$ \\
			\hline
			$\beta_1$   & LSTM units                   & LSTM       & $2^{k_6}$ where $k_6=6, 7, 8$ \\
			\hline
			$\beta_2$   & Learning rate                & LSTM       & $10^{k_7}$ where $k_7=-2,-3,-4$ \\
			\hline
			$\beta_3$     & Batch Size                   & CNN, LSTM  & $2^{k_8}$ where $k_8=6, 7, 8$\\
			\hline
			$\beta_4$     & No. of epochs                & CNN, LSTM  & 10 to 250 in steps of 20 \\ 
			\hline
			\hline
		\end{tabular}
	\end{center}
	\label{table:class-hyper-parameter}
\end{table*}

\subsection{Cross-validation}
\label{sec:crossvalidation}
\begin{figure}
	\centerline{\includegraphics[width=0.5\textwidth]{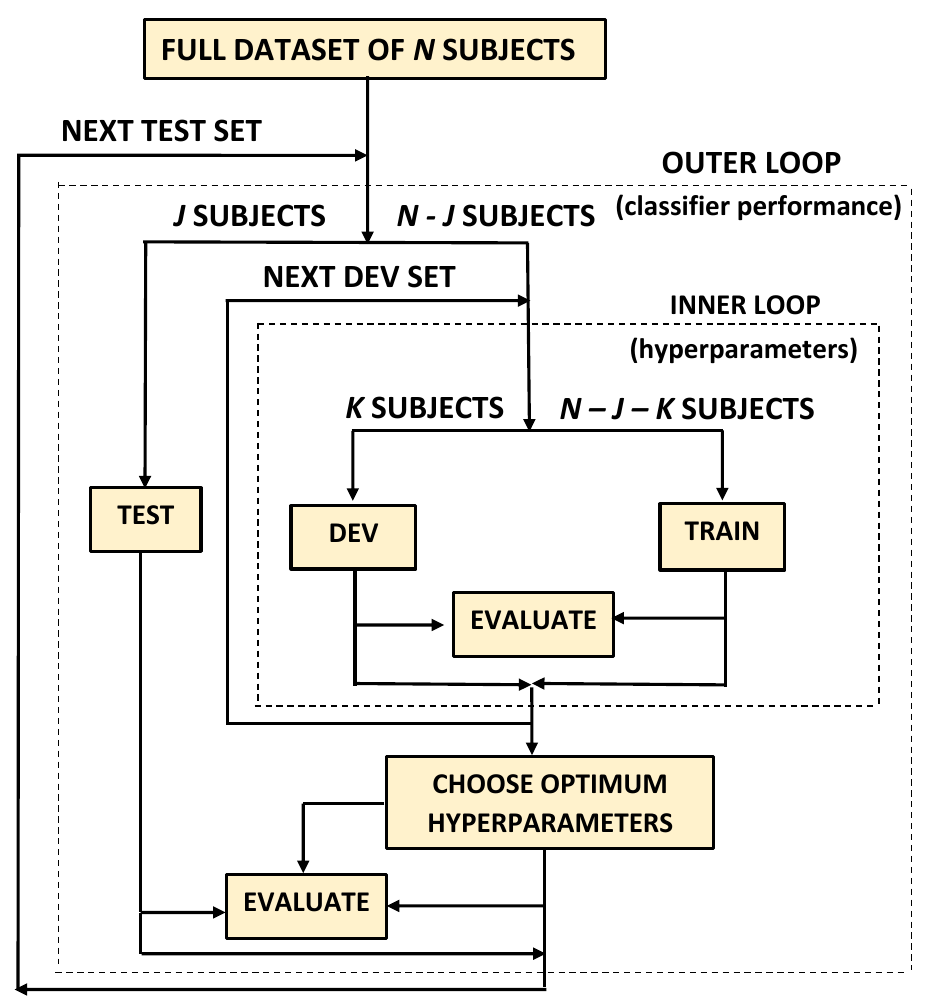}}
	\caption{\textbf{Leave p-out cross-validation}, used to train and evaluate the classifiers. The development set (DEV) consisting $K$ subjects has been used to optimise the hyperparameters while training on the TRAIN set, consisted of $N-J-K$ subjects. The final evaluation of the classifiers in terms of the AUC occurs on the TEST set, consisting $J$ subjects. }
	\label{fig:lpo-k-fold}
\end{figure}

All our classifiers have been trained and evaluated by using a nested leave-$p$-out cross-validation scheme, as shown in Figure~\ref{fig:lpo-k-fold}~\cite{liu2019leave}. 
Since only the Coswara dataset was used for training and parameter optimisation, $N = 1171$ in Figure~\ref{fig:lpo-k-fold}. 
We have set the train and test split as $4:1$; as this ratio has been used effectively in medical classification tasks \cite{racz2021effect}. Thus, $J = 234$ and $K = 187$ in our experiments. 

The figure shows that, in an outer loop, $J$ subjects are removed from the complete set of $N$ subjects to be used for later independent testing.
Then, a further $K$ subjects are removed from the remaining $N-J$ subjects to serve as a development set to optimise the hyperparameters listed in Table~\ref{table:class-hyper-parameter}.
The inner loop considers all such sets of $K$ subjects, and the optimal hyperparameters are chosen on the basis of all these partitions.
The resulting optimal hyperparameters are used to train a final system on all $N-J$ subjects which is evaluated on the test set consisting of $J$ subjects. 
If the $N-J$ subjects in the training portion contain $C_1$ COVID-19 positive and $C_2$ COVID-19 negative coughs, then ($C_2 - C_1$) synthetic COVID-19 positive coughs are created by using SMOTE. 
AUC has always been the optimisation criterion in this cross-validation. 
This entire procedure is repeated for all possible non-overlapping test sets in the outer loop. 
The final performance is evaluated by calculating and averaging AUC over these outer loops. 

This cross-validation procedure makes the best use of our small dataset by allowing all subjects to be used for both training and testing purposes while ensuring unbiased hyperparameter optimisation and a strict per-subject separation between cross-validation folds.

\subsection{Classifier Evaluation}
Receiver operating characteristic (ROC) curves were calculated within the inner and outer loops shown in Figure~\ref{fig:lpo-k-fold}. 
The area under the ROC curve (AUC) indicates how well the classifier has performed over a range of decision thresholds \cite{fawcett2006introduction}. 
From these ROC curves, the decision that achieves an equal error rate ($\gamma_{EE}$) was computed. 
This is the threshold for which the difference between the classifier's true positive rate (TPR) and false positive rate (FPR) is minimised. 

We note the mean per-frame probability that a cough is from a COVID-19 positive subject by $\hat{P}$:
\begin{equation}
	\hat{P} = \frac{\sum\limits_{i=1}^{K} P(Y = 1|X_i, \theta)}{K} 
	\label{eq:P-hat}
\end{equation} 
where $K$ indicates the number of frames in the cough and $P(Y = 1|X_i, \theta)$ is the output of the classifier for feature vector $X_i$ and parameters $\theta$ for the $i^{th}$ frame. 
Now we define the indicator variable $C$ as:
\begin{equation}
	C = 
	\begin{cases}
		1 & \text{if } \hat{P}\geq \gamma_{EE}\\
		0              & \text{otherwise}
	\end{cases}
	\label{eq:C-sum}
\end{equation}

We then define two COVID-19 index scores ($COVID\_I_1$ and $COVID\_I_2$) in Equations~\ref{eq:COVIDI-1} and~\ref{eq:COVIDI-2} respectively. 

\begin{equation}
	COVID\_I_1 = \frac{\sum\limits_{i=1}^{N_1} C}{N_1} 
	\label{eq:COVIDI-1}
\end{equation}

\begin{equation}
	COVID\_I_2 = \frac{\sum\limits_{i=1}^{N_2} P(Y = 1|X_i)}{N_2} 
	\label{eq:COVIDI-2}
\end{equation}

In Equation~\ref{eq:COVIDI-1}, $N_1$ is the number of coughs from the subject in the recording while in Equation~\ref{eq:COVIDI-2}, $N_2$ indicates the total number of frames of cough audio gathered from the subject.
Hence Equation~\ref{eq:P-hat} computes a per-cough average probability while Equation~\ref{eq:COVIDI-2} computes a per-frame average probability. For the Coswara dataset, $N_1 = 1$. 

The COVID-19 index scores, given by Equations~\ref{eq:COVIDI-1} and~~\ref{eq:COVIDI-2}, can both be used to make classification decisions.
We have found that for some classifier architectures one will lead to better performance than the other.
Therefore, we have made the choice of the scoring function an additional hyperparameter to be optimised during cross-validation.


We have calculated the specificity and sensitivity from these predicted values and then compared them with the actual values and finally calculated the AUC and used it as a method of evaluation. 
The mean specificity, sensitivity, accuracy and AUC along with the optimal hyperparameters for each classifier are shown in Table \ref{table:COSWARA-results} and \ref{table:SARCOS-results}.

\section{Results}

\subsection{Coswara dataset}

Classification performance for the Coswara dataset is shown in Table~\ref{table:COSWARA-results}.
The Coswara results are the average specificity, sensitivity, accuracy and AUC along with its standard deviation calculated over the outer loop test-sets during cross-validation. 
These tables also show the values of the hyperparameters which produce the highest AUC during cross-validation.

\begin{table*}[!h]
	\setlength{\tabcolsep}{4pt} 
	\centering
	\caption{\textbf{Classifier performance when training and evaluating on the Coswara dataset.} The best two classifiers along with their feature extraction and optimal classifier hyperparameters are mentioned. The area under the ROC curve (AUC) has been the optimisation criterion during cross-validation. The mean specificity (spec), sensitivity (sens), accuracy (ACC) and standard deviation of AUC ($\sigma_{AUC}$) are also shown. The best performance is achieved by the Resnet50. }
	\begin{tabular}{ c | c | c | c | c | c | c | c }
		\hline
		\hline
		{\multirow{2}{*}{\textbf{Classifier}}} & \textbf{Best Feature} & \textbf{Optimal Classifier Hyperparameters} & \multicolumn{4}{c}{\textbf{Performance}} \\
		
		\cline{4-8}
		
		& \textbf{Hyperparameters} & \textbf{(Optimised inside nested cross-validation)} & \textbf{Spec} & \textbf{Sens} & \textbf{ACC} & \textbf{AUC} & \textbf{$\sigma_{AUC}$} \\
		
		\hline
		\hline
		LR & $\mathcal{M}$=13, $\mathcal{F}$=1024, $\mathcal{S}$=120 & $\nu_1=10^{-4}$, $\nu_2=0.25$, $\nu_3=0.75$ & 57\% & 94\% & 75.70\% & 0.736 & 0.057 \\
		\hline
		LR & $\mathcal{M}$=26, $\mathcal{F}$=1024, $\mathcal{S}$=70 & $\nu_1=10^{-2}$, $\nu_2=0.45$, $\nu_3=0.55$ & 59\% & 74\% & 66.32\% & 0.729 & 0.049 \\
		
		\hline
		KNN & $\mathcal{M}$=26, $\mathcal{F}$=2048, $\mathcal{S}$=100 & $\xi_1=70$, $\xi_2=20$ & 65\% & 83\% & 74.70\% & 0.781 & 0.041 \\
		\hline
		KNN & $\mathcal{M}$=26, $\mathcal{F}$=1024, $\mathcal{S}$=70 & $\xi_1=60$, $\xi_2=25$ & 64\% & 81\% & 73.80\% & 0.776 & 0.039 \\
		
		\hline
		SVM & $\mathcal{M}$=39, $\mathcal{F}$=2048, $\mathcal{S}$=100 & $\zeta_1=10^{-2}$, $\zeta_2=10^{-3}$ & 74\% & 71\% & 72.28\% & 0.815 & 0.046 \\
		\hline
		SVM & $\mathcal{M}$=26, $\mathcal{F}$=1024, $\mathcal{S}$=50 & $\zeta_1=10^{-4}$, $\zeta_2=10^{2}$ & 74\% & 74\% & 73.91\% & 0.804 & 0.051 \\
		
		\hline
		MLP & $\mathcal{M}$=26, $\mathcal{F}$=2048, $\mathcal{S}$=100 & $\eta_1=40$, $\eta_2=10^{-3}$, $\eta_3=0.4$ & 87\% & 88\% & 87.51\% & 0.897 & 0.033 \\
		\hline
		MLP & $\mathcal{M}$=13, $\mathcal{F}$=1024, $\mathcal{S}$=100 & $\eta_1=60$, $\eta_2=10^{-1}$, $\eta_3=0.55$ & 84\% & 68\% & 76.02\% & 0.833 & 0.041 \\
		
		\hline
		CNN & $\mathcal{M}$=26, $\mathcal{F}$=1024, $\mathcal{S}$=70 & $\alpha_1$=48, $\alpha_2$=2, $\alpha_3$=0.3, $\alpha_4$=16, $\beta_3$=128, $\beta_4$=130 & 99\% & 90\% & 94.57\% & 0.953 & 0.039 \\
		\hline
		CNN & $\mathcal{M}$=39, $\mathcal{F}$=1024, $\mathcal{S}$=50 & $\alpha_1$=96, $\alpha_2$=2, $\alpha_3$=0.1, $\alpha_4$=16, $\beta_3$=256, $\beta_4$=170 & 98\% & 90\% & 94.35\% & 0.950 & 0.039 \\
		
		\hline
		LSTM & $\mathcal{M}$=13, $\mathcal{F}$=2048, $\mathcal{S}$=70 & $\beta_1$=128, $\beta_2$=$10^{-3}$, $\alpha_3$=0.3, $\alpha_4$=32, $\beta_3$=256, $\beta_4$=150 & 97\% & 91\% & 94.02\% & 0.942 & 0.043 \\
		\hline
		LSTM & $\mathcal{M}$=26, $\mathcal{F}$=2048, $\mathcal{S}$=100 & $\beta_1$=256, $\beta_2$=$10^{-2}$, $\alpha_3$=0.3, $\alpha_4$=16, $\beta_3$=256, $\beta_4$=110 & 97\% & 90\% & 93.65\% & 0.931 & 0.041 \\
		
		\hline
		\textit{Resnet50} & \textit{$\mathcal{M}$=39, $\mathcal{F}$=1024, $\mathcal{S}$=50} & \textit{Default Resnet50 (Table 1 in \cite{he2016deep})} & \textit{98\%} & \textit{93\%} & \textit{95.33\%} & \textit{0.976} & \textit{0.018} \\
		\hline
		Resnet50 & $\mathcal{M}$=26, $\mathcal{F}$=1024, $\mathcal{S}$=70 & Default Resnet50 (Table 1 in \cite{he2016deep}) & 98\% & 93\% & 95.01\% & 0.963 & 0.011 \\
		
		\hline
		\hline
	\end{tabular}
	\label{table:COSWARA-results}
\end{table*}

\begin{figure}[h!]
	\centerline{\includegraphics[width=0.5\textwidth]{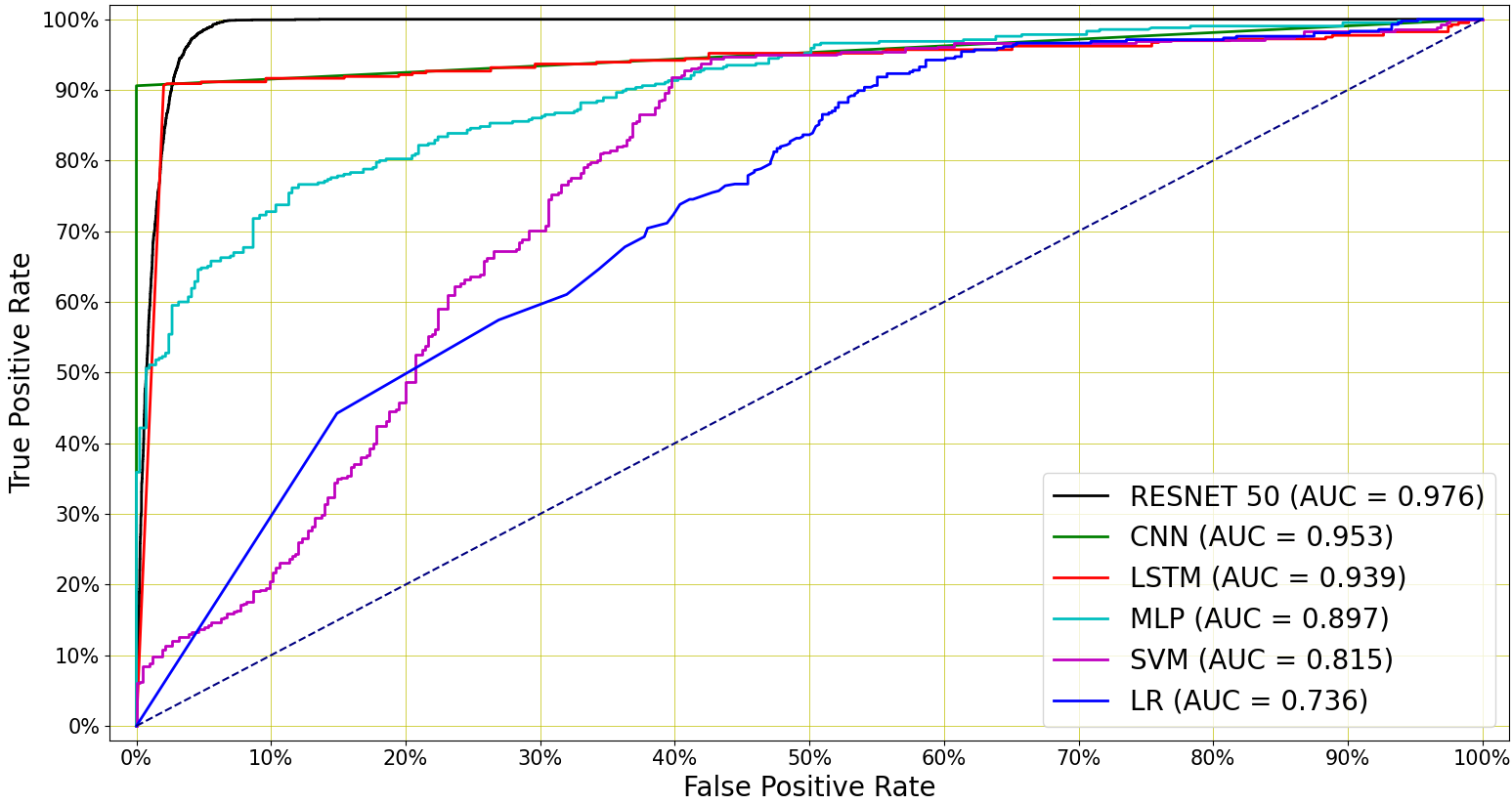}}
	\caption{\textbf{Mean ROC curves for the classifiers trained and evaluated on the Coswara dataset:} The highest AUC of 0.98 was achieved by the Resnet50, while the LR classifier has the lowest AUC of 0.74.}
	\label{fig:mean-ROC-Coswara}
\end{figure}

Table \ref{table:COSWARA-results} shows that all seven classifiers can classify COVID-19 coughs and the Resnet50 classifier exhibits the best performance, with an AUC of 0.976 when using a 120-dimensional feature matrix consisting of 39 MFCCs with appended velocity and acceleration extracted from frames that are 1024 samples long and when grouping the coughs into 50 segments. 
The corresponding accuracy is 95.3\% with sensitivity 93\% and specificity 98\%. 
The CNN and LSTM classifiers also exhibited good performance, with AUCs of 0.953 and 0.942 respectively, thus comfortably outperforming the MLP, which achieved an AUC of 0.897.
The optimised LR and SVM classifiers showed substantially weaker performance, with AUCs of 0.736 and 0.815 respectively. 
Table \ref{table:COSWARA-results} also shows that DNN classifiers exhibit lower standard deviation across the folds than other classifiers. 
This suggests that DNN classifiers are also prone to perform better on new datasets without further hyperparameter optimisation.

The mean ROC curves for the optimised classifier of each architecture are shown in Figure~\ref{fig:mean-ROC-Coswara}. 
We see that LSTM, CNN and Resnet50 classifiers achieve better performance than the remaining architectures at most operating points.
Furthermore, the figure confirms that the Resnet50 architecture also in most cases achieved better classification performance than the CNN and LSTM.
There appears to be a small region of the curve where the CNN outperforms the Resnet50 classifier, but this will need to be verified by future further experimentation with a larger dataset.

We also see from Table~\ref{table:COSWARA-results} that using a larger number of MFCCs consistently leads to improved performance.
Since the spectral resolution used to compute the 39-dimensional MFCCs surpasses that of the human auditory system, we conclude that the classifiers are using information not generally perceivable to the human listeners. 
We have come to similar conclusions in previous work considering the classification of coughing sounds due to tuberculosis \cite{botha2018detection}.

\subsection{Sarcos dataset}
Classification performance for the Sarcos dataset is shown in Table~\ref{table:SARCOS-results}. 
Here the CNN, LSTM and Resnet50 classifiers trained on the Coswara dataset (as shown in Table~\ref{table:COSWARA-results}) were tested on the 44 subjects in Sarcos dataset. 
No further hyperparameter optimisation was performed and hence Table~\ref{table:SARCOS-results} simply notes the same hyperparameters presented in Table \ref{table:COSWARA-results}. 
We see that performance has in all cases deteriorated relative to the better-matched Coswara dataset. 
The best performance was achieved by the LSTM classifier, which achieved an AUC of 0.779. 
In the next section, we improve this classifier by applying feature selection. 

\begin{table*}[!ht]
	\setlength{\tabcolsep}{4pt} 
	\centering
	\caption{\textbf{Classifier performance when training on the Coswara dataset and evaluating on the Sarcos dataset.} The best performance was achieved by the LSTM classifier, and further improvements were achieved by applying SFS.}
	\begin{tabular}{ c | c | c | c | c | c | c }
		\hline
		\hline
		{\multirow{2}{*}{\textbf{Classifier}}} & \textbf{Best Feature} & \textbf{Optimal Classifier Hyperparameters} & \multicolumn{4}{c}{\textbf{Performance}} \\
		
		\cline{4-7}
		
		& \textbf{Hyperparameters} & \textbf{(trained on Coswara dataset in Table \ref{table:COSWARA-results})} & \textbf{Spec} & \textbf{Sens} & \textbf{ACC} & \textbf{AUC} \\
		
		\hline
		\hline
		CNN & $\mathcal{M}$=26, $\mathcal{F}$=1024, $\mathcal{S}$=70 & $\alpha_1$=48, $\alpha_2$=2, $\alpha_3$=0.3, $\alpha_4$=16, $\beta_3$=128, $\beta_4$=130 & 61\% & 85\% & 73.02\% & 0.755 \\

		\hline
		LSTM & $\mathcal{M}$=13, $\mathcal{F}$=2048, $\mathcal{S}$=70 & $\beta_1$=128, $\beta_2$=$10^{-3}$, $\alpha_3$=0.3, $\alpha_4$=32, $\beta_3$=256, $\beta_4$=150 & 73\% & 75\% & 73.78\% & 0.779 \\
		
		\hline
		Resnet50 & $\mathcal{M}$=39, $\mathcal{F}$=1024, $\mathcal{S}$=50 & Default Resnet50 (Table 1 in \cite{he2016deep}) & 57\% & 93\% & 74.58\% & 0.742 \\
		
		\hline
		\textit{LSTM + SFS} & \textit{$\mathcal{M}$=13, $\mathcal{F}$=2048, $\mathcal{S}$=70} & \textit{$\beta_1$=128, $\beta_2$=$10^{-3}$, $\alpha_3$=0.3, $\alpha_4$=32, $\beta_3$=256, $\beta_4$=150} & \textit{96\%} & \textit{91\%} & \textit{92.91\%} & \textit{0.938} \\
		
		\hline
		\hline
	\end{tabular}
	\label{table:SARCOS-results}
\end{table*}

\subsubsection{Feature Selection}

As an additional experiment, SFS has been applied to the best-performing system in Table \ref{table:SARCOS-results}, the LSTM. 
SFS is a greedy selection method for the individual feature dimensions that contribute the most towards the classifier performance \cite{devijver1982pattern}.

\begin{figure}[h!]
	\centerline{\includegraphics[width=0.5\textwidth]{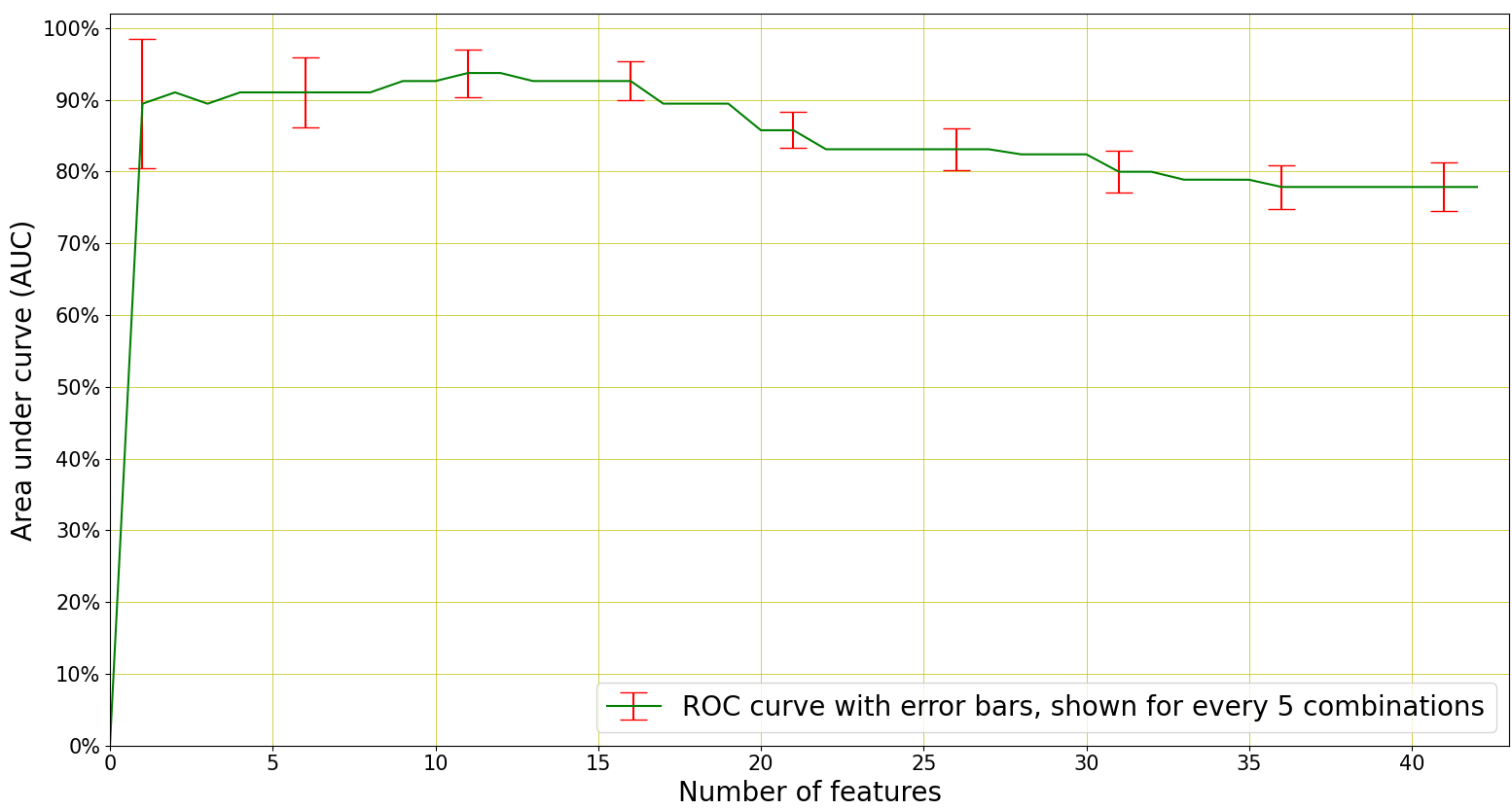}}
	\caption{\textbf{Sequential Forward Selection}, when applied to a feature matrix composed of 13 MFCCs with appended velocity ($\Delta$) and acceleration ($\Delta \Delta$), log frame energies, ZCR and kurtosis (Equation \ref{eq:seg}). Peak performance is observed after selecting the best 13 features. }
	\label{fig:best-SFS}
\end{figure}

\begin{figure}[h!]
	\centerline{\includegraphics[width=0.5\textwidth]{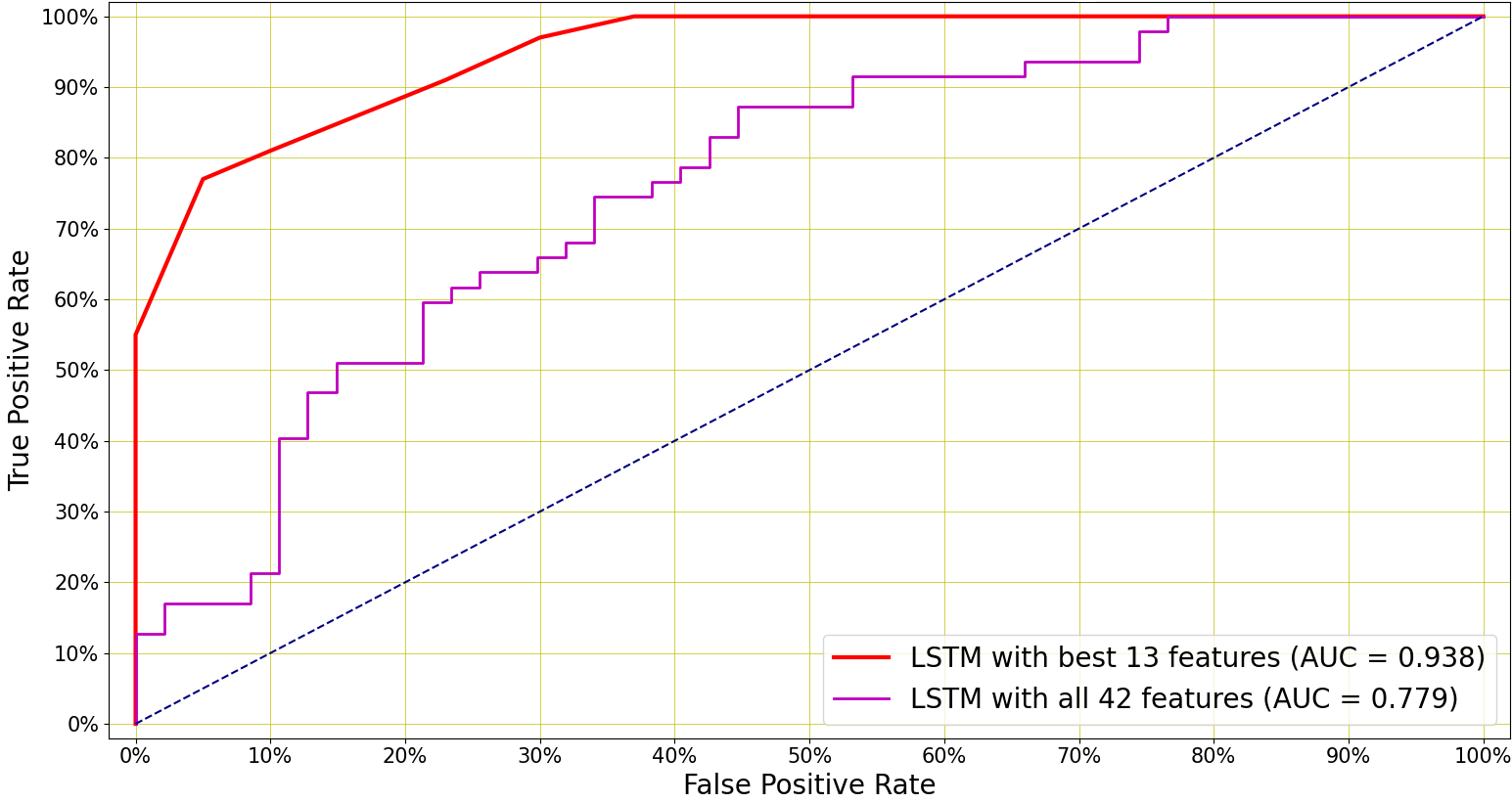}}
	\caption{\textbf{Mean ROC curve for the best performed LSTM classifier trained on Coswara dataset and evaluated on Sarcos dataset:} AUC of 0.78 has been achieved while using all 42 features. After applying SFS and selecting the best 13 features, the AUC has been improved to 0.94.}
	\label{fig:mean-ROC-SARCOS-best}
\end{figure}

The feature selection hyperparameters in these experiments were 13 MFCCs extracted from 2048 samples (i.e. 0.46 sec) long frames while coughs were grouped into 70 segments. 
Thus, SFS could select from a total of 42 features: MFCCs along with their velocity ($\Delta$) and accelerations ($\Delta \Delta$), log frame energy, ZCR and Kurtosis (Equation \ref{eq:seg}). 
After performing SFS to the LSTM classifier, a peak AUC of 0.938 was observed on the Sarcos dataset when using the best 13 features among those 42, as shown in Figure \ref{fig:best-SFS} and Table \ref{table:SARCOS-results}. 
These 13 selected features led to an improvement of AUC from 0.779 to 0.938 (Figure \ref{fig:mean-ROC-SARCOS-best}) and they include MFCCs ranging from 3 to 12 along with their velocity ($\Delta$) and acceleration ($\Delta \Delta$), suggesting all dimensions of feature matrix carry equally-important COVID-19 signatures. 

\section{Conclusion and Future Work}

We have developed COVID-19 cough classifiers using smartphone audio recordings and seven machine learning architectures. 
To train and evaluate these classifiers, we have used two datasets. 
The first, larger, dataset is publicly available and contains data from 1171 subjects (92 COVID-19 positive and 1079 healthy) residing on all five continents except Africa. 
The smaller second dataset contains recordings from 18 COVID-19 positive and 26 COVID-19 negative subjects, 75\% of whom reside in South Africa.  
Thus, together the two datasets include data from subjects residing on all six continents. 
After pre-processing the cough audio recordings, we have found that the COVID-19 positive coughs are 15\%-20\% shorter than non-COVID coughs.
Then we have extracted MFCCs, log frame energy, ZCR and kurtosis features from the cough audio using a special feature extraction technique which preserves the time-domain patterns and then trained and evaluated those seven classifiers using the nested leave-$p$-out cross-validation. 
Our best-performing classifier is the Resnet50 architecture and is able to discriminate between COVID-19 coughs and healthy coughs with an AUC of 0.98 on the Coswara dataset. 
These results outperform the baseline result of the AUC of 0.7 in \cite{muguli2021dicova}. 
When testing on the Sarcos dataset, the LSTM model trained on the Coswara dataset exhibit the best performance, discriminating COVID-19 positive coughs from COVID-19 negative coughs with an AUC of 0.94 while using the best 13 features determined by sequential forward selection (SFS). 
Furthermore, since better performance is achieved using a larger number of MFCCs than is required to mimic the human auditory system, we also conclude that at least some of the information used by the classifiers to discriminate the COVID-19 coughs and the non-COVID coughs may not be perceivable to the human ear.

Although the systems we describe require more stringent validation on a larger dataset, the results we have presented are very promising and indicate that COVID-19 screening based on automatic classification of coughing sounds is viable.
Since the data has been captured on smartphones, and since the classifier can in principle also be implemented on such device, such cough classification is cost-efficient, easy to apply and deploy.
Furthermore, it could be applied remotely, thus avoiding contact with medical personnel.

In ongoing work, we are continuing to enlarge our dataset and to apply transfer learning in order take advantage of the other larger datasets. 
We are also beginning to consider the best means of implementing the classifier on a readily-available consumer smartphone.

\section*{Acknowledgements}
This project was funded by the South African Medical Research Council (SAMRC) through its Division of Research Capacity Development under the SAMRC Intramural Postdoctoral programme, the South African National Treasury, as well as an EDCTP2 programme supported by the European Union (TMA2017CDF-1885). 
We would like to thank the South African Centre for High Performance Computing (CHPC) for providing computational resources on their Lengau cluster for this research, and gratefully acknowledge the support of Telcom South Africa. 
We would also like to thank Jordan Govendar and Rafeeq du Toit for their invaluable assistance in the Sarcos data collection.

The content and findings reported are the sole deduction, view and responsibility of the researchers and do not reflect the official position and sentiments of the SAMRC, EDCTP2, European Union or the funders.



\bibliography{IEEEabrv, reference}
\bibliographystyle{IEEEtran}





\begin{IEEEbiography}[{\includegraphics[width=1in,height=1.25in,clip,keepaspectratio]{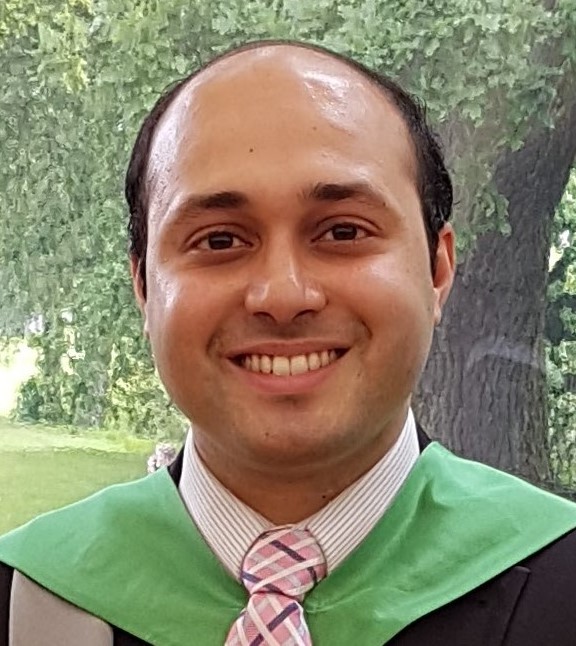}}]{Madhurananda Pahar} received his BSc in Mathematics from the University of Calcutta, India, and his MSc in Computing for Financial Markets followed by his PhD in Computational Neuroscience from the University of Stirling, Scotland. Currently, he is working as a post-doctoral fellow at the University of Stellenbosch, South Africa. His research interests are in machine learning and signal processing for audio signals and smart sensors in bio-medicine such as the detection and classification of TB and COVID-19 coughs in a real-world environment.

\end{IEEEbiography}

\begin{IEEEbiography}[{\includegraphics[width=1in,height=1.25in,clip,keepaspectratio]{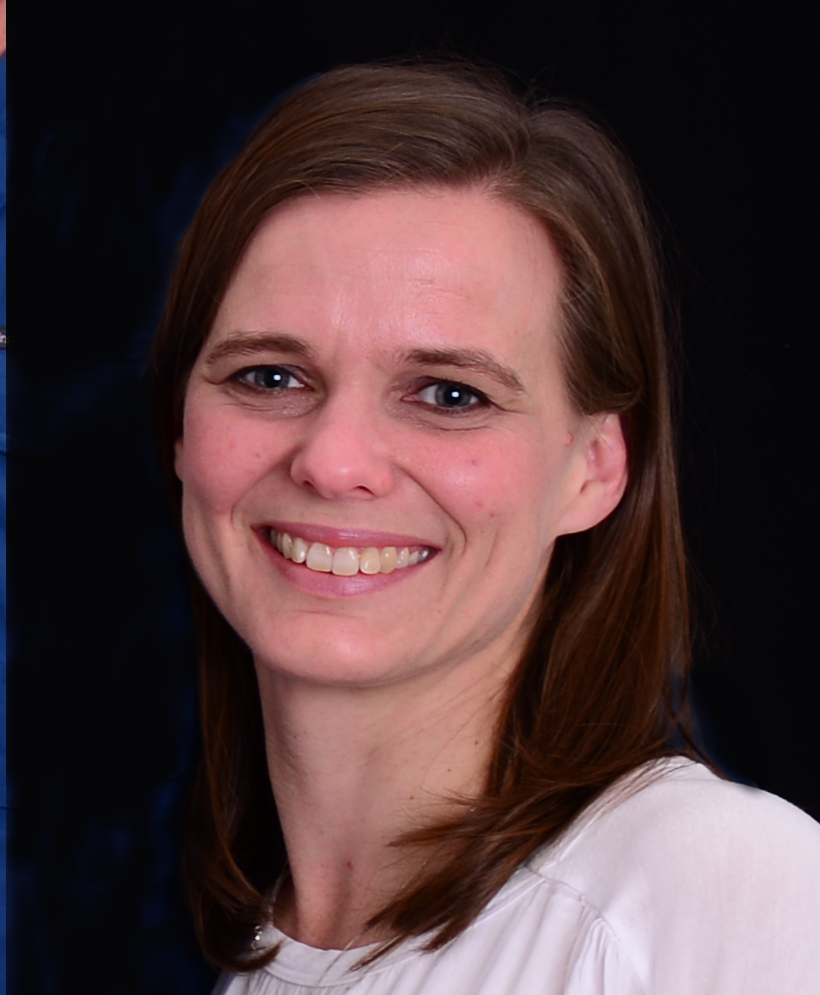}}]{Marisa Klopper} is a researcher at the Division of Molecular Biology and Human Genetics of Stellenbosch University, South Africa. She holds a PhD in Molecular Biology from Stellenbosch University and her research interest is in TB and drug-resistant TB diagnosis, epidemiology and physiology. She has been involved in cough classification for the last 6 years, with application to TB and more recently COVID-19. 
\end{IEEEbiography}

\begin{IEEEbiography}[{\includegraphics[width=1in,height=1.25in,clip,keepaspectratio]{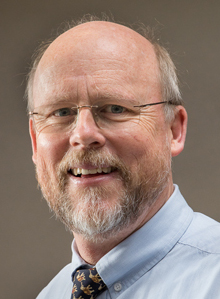}}]{Robin Warren} is the Unit Director of the South African Medical Research Council’s Centre for Tuberculosis Research and Distinguished Professor at Stellenbosch University. He has a B2 rating by the National Research Council (NRF) and is a core member of the DSI-NRF Centre of Excellence for Biomedical Tuberculosis Research and head the TB Genomics research thrust. He has published over 320 papers in the field of TB and have an average H-index (Scopus, Web of Science and Google Scholar) of 65. 
\end{IEEEbiography}

\begin{IEEEbiography}[{\includegraphics[width=1in,height=1.25in,clip,keepaspectratio]{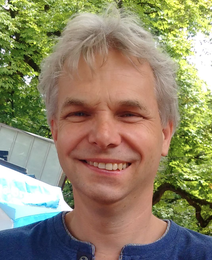}}]{Thomas Niesler} obtained the B.Eng (1991) and M.Eng (1993) degrees in Electronic Engineering from the University of Stellenbosch, South Africa and a Ph.D. from the University of Cambridge, England, in 1998. He joined the Department of Engineering, University of Cambridge, as a lecturer in 1998 and subsequently the Department of Electrical and Electronic Engineering, University of Stellenbosch, in 2000, where he has been Professor since 2012. 
His research interests lie in the areas of signal processing, pattern recognition and machine learning.
\end{IEEEbiography}

\end{document}